\documentstyle{article}
\begin{document}
\title{Controversies in the History of the Radiation Reaction problem
in General Relativity}
\author{Daniel Kennefick}
\date{}
\maketitle
\section{Introduction}
Beginning in the early 1950s, experts in
the theory of general relativity debated vigorously 
whether the theory predicted the emission
of gravitational radiation from
binary star systems. For a
time, doubts also arose  
on whether gravitational waves could carry any energy.
Since radiation phenomena have played a key role in the development of
20th century field theories, it is 
the main purpose of this paper 
to examine the
reasons for the growth of scepticism regarding radiation in the case of
the gravitational field.
Although the focus is on
the period from the mid-1930s 
to about 1960, when the modern study
of gravitational waves was developing,
some attention is also paid to the 
more recent and unexpected emergence of experimental
data on gravitational waves which considerably 
sharpened the debate on certain controversial
aspects of the theory of gravity waves. 
I analyze the use of the earlier
history as a rhetorical device in review papers written by protagonists of the
``quadrupole formula controversy'' in the late 1970s and early 1980s.
I argue that relativists
displayed a lively interest in the 
historical background to the problem
and exploited their 
knowledge of the literature to justify
their own work and their 
assessment of the contemporary state of the subject.
This illuminates 
the role of a scientific field's sense of its
own history as a mediator in scientific controversy.$^{1}$

\section{The Einstein-Rosen Paper}
In a letter to to his friend Max Born, probably written sometime
during 1936, Albert Einstein reported 

\begin{quotation}
Together with a young collaborator, I arrived at the interesting result 
that gravitational waves do not exist, though they had been assumed a 
certainty to the first approximation. This shows that the non-linear general 
relativistic field equations can tell us more or, rather, limit us more than 
we have believed up to now.$^2$ (Born 1971, p. 125)
\end{quotation}

The young
collaborator was Nathan Rosen, with whom Einstein had been working for
some time, producing papers on several topics.
They had submitted a paper to the {\it Physical Review}
based on the work referred to in Einstein's letter to Born
under the title ``Do Gravitational Waves Exist?''$^3$ 
and the answer they proposed to give, as the letter states, was no. It is 
remarkable that at this stage in his career, Einstein was prepared
to believe that gravitational waves did not exist, 
all the more so because he had made them one of the first predictions
of his theory of general relativity.
In his autobiography Leopold Infeld, who arrived 
in Princeton in 1936
to begin an important collaboration with Einstein,
described
his surprise 
on hearing of the result (Infeld 1941, pg. 239). 
Despite his initial scepticism, Infeld soon allowed himself
to be convinced by Einstein's arguments, and even came up with his own
version of the proof, which reinforced his belief in the result
(Infeld 1941, pg. 243). However, not everyone was 
so easily convinced. When Einstein sent the paper
to the {\it Physical Review} for publication, it was returned to him with a
critical referee's report (EA 19-090), accompanied by 
the editor's mild request that 
he ``would be glad to have your reaction to the various comments
and criticisms the referee has made.''
(John T. Tate to Einstein July 23, 1936, EA 19-088). Instead, Einstein wrote
back in high dudgeon, withdrawing the paper, and dismissing out of hand
the referee's comments (Einstein to Tate July 27, 1936, EA 19-086): 

\begin{quotation}
Dear Sir,

           We (Mr. Rosen and I) had sent you our manuscript for 
\underline{publication} and had not authorized you to show it to
specialists before it is printed. I see no reason to address the
- in any case erroneous - comments of your anonymous expert. On the
basis of this incident I prefer to publish the paper elsewhere.

                                                      respectfully,

P.S. Mr. Rosen, who has left for the Soviet Union, has authorized 
me to represent him in this matter.$^4$
\end{quotation}

To this Tate replied that he regretted Einstein's decision to withdraw
the paper, but stated that he would not set aside the journal's review 
procedure.
In particular, he 
``could not accept for publication
in THE PHYSICAL REVIEW a paper which the author was unwilling I should
show to our Editorial Board before publication.'' (Tate to Einstein
July 30, 1936, EA 19-089). Einstein must have continued in his dislike of
the Review's editorial policy (which in fairness may have been
unfamiliar to him, the practice of German journals being
less fastidious$^{5}$), for he never published there
again.$^{6}$
The paper with Rosen was, however, subsequently accepted
for publication by the Journal of the Franklin Institute in Philadelphia.$^7$

What had led Einstein to the conclusion which so surprised Infeld? He and
Rosen had set out to find an exact solution to the field equations of 
general relativity 
which described plane gravitational waves, and had found themselves
unable to do so without introducing singularities 
into the components of the metric describing the wave. As a result,
they felt they could show that no regular periodic wavelike 
solutions to the equations
were possible (Rosen 1937 and 1955). 
However, in July of 1936, the relativist 
Howard Percy Robertson returned
to Princeton from a sabbatical year in Pasadena and
subsequently struck up a friendship with the newly arrived 
Infeld. He told Infeld that he 
did not believe Einstein's
result, and his scepticism was much less shakeable. Certain
that the result was incorrect, he went over Infeld's version of the argument
with him, and they discovered an error
(Infeld 1941, pg. 241). When this was communicated
to Einstein, he quickly concurred and made changes in proof to the paper
which was then with the Franklin journal's publisher (Infeld 1941, pg. 244
and letter, Einstein to editor of the 
Franklin Journal November 13, 1936, EA 20-217).$^{8}$ 

Although a footnote attached to the published version acknowledges Robertson's
help, it does not indicate its nature (Einstein and Rosen 1937). 
However, it appears that
his chief contribution was to observe that the singularity could be avoided
by constructing a cylindrical wave solution. In this way the offending
singularity would 
be relegated to the infinitely long central symmetry 
axis of the wave, where it was
less objectionable, being identifiable with a material 
source (Rosen 1955).  
In view of this, Einstein might have been better advised
not to dismiss the referee's report so hastily, 
as the anonymous reviewer also observed that, by casting the Einstein-Rosen
metric in cylindrical co-ordinates the apparent difficulty with the metric was
removed, and it was easily seen to be describing cylindrical waves 
(Referee's report, EA 19-090, pgs. 2,3,5).$^9$ That Robertson was familiar
with the referee's criticisms 
is shown by his letter to Tate of February 18, 1937 (Caltech archives,
Robertson papers, folder 14.6)
in which he says

\begin{quotation}
You neglected to keep me informed on the paper submitted last summer by
your most distinguished contributor. But I shall nevertheless let you in
on the subsequent history. It was sent (without even the correction of one
or two numerical slips pointed out by your referee) to another journal,
and when it came back in galley proofs was completely revised because I
had been able to convince him in the meantime that it proved the opposite
of what he thought.

You might be interested in looking up an article in the Journal of the
Franklin Institute, January 1937, p. 43, and comparing the conclusions
reached with your referee's criticisms.
\end{quotation}

This suggests that, in spite of himself, Einstein did benefit from the
referee's advice in the end, by a very circuitous route.

In fact the cylindrical wave solution presented in the revised paper 
had been previously published
by the Austrian physicist Guido Beck in 1925, but his paper has been largely
overlooked since. In a 1926 paper by Baldwin and Jeffrey, and in the 
referee's report on Einstein's paper, there was discussion of the fact   
that singularities in the metric coefficients are unavoidable
when describing plane waves with infinite wave fronts, but although 
there is some distortion in the wave, ``the field
itself is flat'' at infinity, as the referee noted (EA 19-090, pg. 9).
In any case, the Einstein-Rosen paper, as published,
contains no direct reference to any other paper whatever.
Rosen published a paper in 1937 in a Soviet journal,
carrying through what is presumably
the chief argument of the original version of the Einstein-Rosen paper, 
in order to show that
plane gravitational waves were an impossibility 
due to the ineradicability
of singularities in the metric. In the immediate post-war period, other
papers suggested that plane waves were not permitted in General 
Relativity (for example, McVittie 1955).
Felix Pirani and Hermann Bondi 
were both partly motivated by these papers to work on the
problem of gravitational waves.$^{10}$ In the mid-fifties, Ivor
Robinson independently rediscovered the plane wave metric and,
together with Bondi and Pirani, published the seminal work on the subject.
They were familiar with Rosen's paper, and noted that his regularity
conditions for the metric were unnecessarily 
severe by post-war standards. ``In
effect, Rosen did not distinguish sufficiently between co-ordinate
singularities and physical singularities, which could, in principle,
be detected experimentally'' (Bondi, Pirani and Robinson 1959).$^{11}$

\section{Gravitational Radiation since Einstein}

In 1916, in a paper exploring the physical implications of the 
final version of his general theory of relativity,
Einstein proposed the
existence of gravitational radiation as one of its important consequences
(Einstein 1916).
Although both Maxwell and Poincar\'{e} have been cited as
anticipating the idea of gravitational waves (Havas 1979 and Damour 1987a),
Einstein
produced the first concrete description in a relativistic field theory. 
In a subsequent paper of 1918, Einstein corrected some errors in his
previous description of the waves, and went on to calculate
the flux of energy  
carried by the waves far from their source (Einstein 1918). 
Appealing to the principle of conservation of energy, he assigned an
equivalent loss of energy to the source system, an effect 
already familiar 
from electromagnetic theory, nowadays
known variously as ``radiation reaction'',
``back reaction'' or, in cases involving the decay of periodic motion
such as orbital motion, ``radiation damping''. Because Einstein's formula
for the energy emission depended on changes in 
the mass quadrupole moment of
the source, it became known as the quadrupole formula. 
In deriving the formula, Einstein made use of a linearized version
of his field equations both for ease of manipulation and because of its
strong analogy 
to the field equations of electromagnetism. Not surprisingly, therefore,
his quadrupole formula was itself similar in form to the multipole
radiation formulas of electromagnetism, in which field, however, the lowest
order of emission is the dipole.

Einstein was not the first to discuss gravitational radiation
reaction. In 1908, Poincar\'{e} had suggested that planetary orbits 
must slowly lose energy to wave emission in the gravitational field
and indicated that any such effect would be too small to explain the perihelion
shift of Mercury
(Poincar\'{e} 1908). As early
as 1776, Pierre Laplace had
considered the problem of an orbital damping force 
arising from a finite speed of propagation of gravity. His aim was to
discover an explanation for the observed 
decrease of the Moon's orbital period
with respect to ancient eclipse observations
(Laplace 1776). 

In general there are two distinguishable approaches to the back
reaction problem. The first, and generally the simpler is the energy
balance argument used by Einstein in his 1918 paper. 
This approach has
been criticized in principle on several
counts in the context of general relativity, but was an obvious choice
for a first approximation. 
The second approach, more direct but much more complex,
is to iteratively calculate the effect of the source's own field
(changing because of the source's motion), upon the source's motion,
corrections to which can then be reapplied to calculate the field
more accurately.
This iteration is carried through one or more steps
until it is judged that the reaction effects have been calculated
to the desired level of accuracy.
This problem is part of a more general one known as the problem of motion.
Laplace's method, which took into account the deflection of the 
Newtonian central force on an orbiting body 
as a result of the time lag in propagation, was a ``one-step''
calculation of this type.
A key issue in this approach is the fact that the field,
in the case of finite propagation, is ``retarded'', which is to say
that the field experienced at a given point in space, at a given time is
not that produced by the source at that time, but that of the source
at an earlier time, where the difference between the two times is the
time of propagation of the field changes from the source's 
retarded position to the
field point in question.
As Laplace showed,
an orbital decay would be one consequence of introducing retarded
propagation instead of dealing with instantaneous propagation. His 
ultimate conclusion, however,
was that the lunar orbital decay could be explained by other, conservative
gravitational effects. Therefore finite propagation times
had no observable effect in real systems, and the
instantaneous action-at-a-distance hypothesis of the day was justified
(Laplace 1825). $^{12}$

\section{Later work on Radiation}

Arthur Stanley Eddington is associated with the remark that gravitational waves
propagate with ``the speed of thought'' (Eddington 1922). 
Despite the scepticism this implies, Eddington was arguing only that
certain classes of gravity waves, the ``transverse-longitudinal'' 
and ``longitudinal-longitudinal'' waves 
analyzed by Weyl (1921) and Einstein (1918) were
unphysical. As mere coordinate effects they could be propagated
with any velocity desired by the human mind. 
In the linearized theory at least, Eddington could show that
transverse-transverse waves could carry energy, and he 
reproduced Einstein's quadrupole formula while correcting an
erroneous factor of two in Einstein's early version
(Eddington 1922, pg. 279). He noted, at the
same time, that the linearized theory was invalid for sources 
such as binary
stars, in which the system was held together by gravitational forces
(Eddington 1922, pg. 280).
In 1941, the Russian physicists Lev Landau and Evgenii Lifschitz
published a back reaction calculation
which did treat a binary star system, including its gravitational
binding, in the slow-motion weak-field case (Landau and Lifschitz 1951).
Their analysis has been influential, although
some have felt that it took too much for granted,
a problem worsened by the book's terse style.

Although the main topic of the Einstein-Rosen paper
had nothing explicitly to do with the back reaction
problem, it is very noteworthy as the first serious (if abortive) 
attempt to disprove
the existence of gravitational waves. 
In an interesting passage
addressing radiation reaction, 
the published paper suggests that one is not compelled
to the conclusion that waves emitted by a source must damp the source's
motion, if one supposes that any outbound radiant energy is matched by
a second system of incoming waves, impinging on the source. In short,
they observed that the use of half-advanced plus half-retarded potentials
will avoid motion damping in the source system even if the waves exist.
``This leads to an undamped mechanical process which is embedded in a
system of standing waves,'' in the author's words 
(Einstein and Rosen 1937). 
The paper refers cryptically to the work of
Ritz and Tetrode ``in former years'' relating to the question of advanced
versus retarded potentials
(in which the field at time $t$ is that produced by the source
from a {\it future} or a {\it past} position respectively),
and it appears that Einstein often quoted Ritz 
approvingly in this context (Infeld and Plebanski 1960, pg.201). 

Walter Ritz, a Swiss 
contemporary
and friend of Einstein's had complained 
in his criticism of Lorentz's electrodynamics that
advanced potentials
were admitted as solutions of the equations
of electrodynamics just as well as the retarded potentials (Ritz 1908). 
To Ritz, this
defied the principle of causality, since effect preceded cause. Just
as abhorrent to Ritz were combinations of the two potentials, such as
the average of advanced and retarded fields (half-advanced plus half-retarded)
which allowed
``perpetual'' motion because, like the instantaneous interaction, it
produced no motion damping due to back reaction.
Ironically, what Ritz regarded as so damning, Einstein appears
to imply might have a positive virtue, in the context of gravitation.$^{13}$

The Dutch physicist Hugo Tetrode, also an acquaintance of Einstein,
discussed the half-advanced-plus-half-retarded 
potential in a paper of 1922. At the
time this solution
to the classical wave equations seemed a possible explanation
for the failure of 
orbiting atomic electrons to radiate. Furthermore, as Tetrode
pointed out, in the quantum regime, the emission and absorption of
radiation seemed to each depend on the other, rather than emission being
required for absorption, but not the reverse. This suggested to him that
the classical aversion to making absorption a requirement for emission
should be discarded. As he put it, ``The Sun would not shine if it were
alone in the universe'' (Tetrode 1922). 
In their paper, Einstein and Rosen appear to share 
Tetrode's preference for this potential, if not for his full 
action-at-a-distance program.

The story, in any case, is of particular concern to us, because of the 
project upon which Einstein and Infeld now embarked together with
Banesh Hoffman. They wished to develop the post-Newtonian
theory of the problem of motion, an ambitious project involving intensive
calculations (Einstein, Infeld and Hoffman 1938). 
Since the non-linear field equations of relativity are too
complex to be solved exactly for dynamical
systems of masses, approximation schemes are required. 
In general relativity, two different schemes have been commonly employed. 
The post-Newtonian expansion makes corrections to the Newtonian 
motion of the system. 
Since the Newtonian limit is only valid for weak fields and
slow motion, the expansion is in powers of the field strength 
(expansion parameter $(G/c^2)(m/r)$, where $G$ is the gravitational
constant, $c$ the speed of light, and $m$ and $r$ represent internal
masses and distances of the source) and the
source velocities (expansion parameter $v/c$, where $v$ represents
small velocities of the source). An alternative approach 
is to make corrections to the linearized
equations of motion, in an expansion based on powers of the field strength
alone.
Because it was not limited to small velocities, 
the second approach became known as
the fast motion approximation (and since the 1970s as ``post-linear''
or ``post-Minkowski''). For a modern review of approximation methods
in the problem of motion, see Damour (1987b).

The problem of motion had been 
previously tackled by Einstein and others,$^{14}$
but the post-Newtonian Einstein-Infeld-Hoffman (EIH) method, 
was to be one of the more
influential, in a very general way.
Einstein particularly wished to vindicate
his conjecture that in general
relativity the allowed motions of
the particles were completely determined by the field equations
(Einstein and Grommer 1927), in contrast
to other field theories where a separate force law is invoked.

Not long after the work
was successfully 
completed, Infeld, who had with Robertson's help secured a position
at the University of Toronto, put his graduate student Phillip Wallace to
work applying the EIH formalism to the problem of motion in electrodynamics.
In their paper,
as also in the EIH paper itself (where radiation effects were not considered),
we see a preference for the averaged potential, ``half advanced
plus half retarded''. Infeld and Wallace state
that this solution ``does not specify
a privileged direction for the flow of time'' and is besides the simplest 
for their method (Infeld and Wallace 1940). 
They note that this solution does not damp orbital
motion, 
and further state that ``the addition of radiation seems from this
point of view arbitrary'', since one must choose the retarded potential
to obtain it. This viewpoint partly reflects Einstein's own.
The solutions which admit radiation damping are objectionable because
they involve an arbitrary imposition of the arrow of time into field
theories which are otherwise time-symmetric.
Although Ritz had pointed out how this arbitrariness was an unsatisfactory
feature of electrodynamics, his conclusion had been that one must choose
the retarded potential to make any sense of it, until a theory which imposed
it could be found. Einstein however, felt that time asymmetry
had no business in field theories and that its origins lay solely in
probability theory(Einstein and Ritz 1909). His views may have influenced
Infeld, who preferred the half-advanced-plus-half-retarded potential,
with its standing wave solution, as the most natural
choice in the EIH approximation. In the case of the gravitational field, where
the existence of radiation could not be experimentally proven, 
Infeld may have felt there was no compulsion to impose the arrow of
time, as one would in electromagnetism, knowing from experiment
that radiation existed in
that field. 

In the 1970s, Rosen returned to the 
problem of the arrow of time in gravitational
radiation theory, 
in a paper
whose title notably echoed that of his rejected 1936 submission to
Physical Review with Einstein (Rosen 1979). 
In ``Does Gravitational Radiation Exist?'' he
adapted the Wheeler-Feynman absorber theory to gravitation, and concluded
that as the gravitational force interacted much less strongly with matter
than the electromagnetic field, a source system would not undergo
radiation reaction for lack of a sufficiently strong absorber field. In the
Wheeler-Feynman theory it is the field of the absorbers, back-reacting
on the source, which breaks the time symmetry of the source field.
(Wheeler and Feynman, 1945 and 1949). However,
Rosen's arguments do not 
appear completely convincing even to himself, since towards
the end of the paper he retreats to a more Tetrode-like position,
conceding that an absorber (such as a gravity wave detector) 
could presumably act so as to draw energy from
the source at a distance. In any case, his paper did not excite much
debate on the subject.

\section{Post-War work}

The first post-Newtonian attempts to deal with gravitational radiation
reaction via the problem of motion 
had to wait until after the war. 
In 1946 Ning Hu, a Chinese graduate of Caltech, presented
results based
on a scheme inspired by the EIH method
to the Royal Irish Academy in Dublin, reporting an energy loss
disagreeing with the quadrupole formula 
in the case of an equal mass binary system in a circular orbit (Hu 1947).
Shortly before publication, however, he added a note in proof
after finding a calculational error which changed the sign of his result,
giving anti-damping instead of damping. In other words, the system would
gain, rather than 
lose energy as the result of emitting radiation. The binary
would therefore slowly increase, not decrease in radius. 
In Canada, Infeld 
and his student,
Adrian Scheidegger, worked on the problem of gravitational
radiation reaction in the EIH formalism (Infeld and Scheidegger 1951). 
They concluded that
the most natural treatment of the scheme,
employing the standing wave boundary condition, led to a no-radiation-reaction
result. It was possible, they conceded, to find terms at certain large
odd powers of $v/c$ 
which appeared to correspond to 
back-reaction terms, but
they contended that these could always be transformed away by a suitable
choice of co-ordinates. The result, when announced
at an American Physical Society meeting in 1950, ``gave rise to a
considerable flow of discussion'', as Scheidegger put it (Scheidegger 1951).
That same year Infeld left Canada, after
a McCarthyite campaign against him organized in the press and
in parliament, absurdly alleging that he was in possession of atomic secrets.
He returned to his native Poland, while
Scheidegger continued to argue the no-damping position
in North America in his absence, before leaving the field of general relativity
for that of geophysics in the mid-fifties.

In 1955 came two futher contributions. Joshua Goldberg, a student of
Peter Bergmann (who had criticized the Infeld and Scheidegger results),
examined the reaction problem in the EIH formalism (Goldberg 1955). 
His conclusions
were twofold. On the one hand, he denied that the slow motion approach
tended to exclude the possibility of damping (arguing that co-ordinate
transformations which removed some back-reaction terms, would
reintroduce other reaction terms of odd order in $v/c$), but on the other hand,
he determined that it was poorly suited to the back reaction problem,
principally because of the restriction to slow motions of the source.
In fact, it was generally agreed that radiation reaction terms did not
enter into the post-Newtonian equations of motion until terms of order at least $(v/c)^5$
beyond Newtonian order. Since leading order
post-Newtonian effects ($(v/c)^2$ order), such as those obtained
by EIH, were both small and difficult to calculate, the expansion method
seemed unpromising for studying radiation 
in that it had to be pushed to high order to succeed.
A couple of years later Goldberg was introduced to Peter Havas,
a physicist
with experience in the problem of radiation in special relativity, who
shared his interest in developing a fast motion expansion in 
general relativity.
Having each worked on the problem
independently, they began a collaboration based on this approach.$^{15}$

Also in 1955, the Russian physicist
Vladimir Fock treated the orbital damping problem in his book {\it Spacetime
and Gravitation} (Fock 1959). He made use of a slow-motion expansion
which he had developed
independently of EIH, coupled with ``no-ingoing wave'' boundary conditions
in the past of the system. His results were in agreement with those
of Landau and Lifschitz. His work was not translated into English for
four years, and even then wielded little influence in the west, perhaps
because of Fock's unorthodox views on general covariance. He employed
so-called harmonic co-ordinates in his calculations, and claimed a special
status for them in physical theory. His views in this regard were 
vigorously opposed by Infeld and most other relativists then and since.
Furthermore, Fock himself regarded his back-reaction result as merely
demonstrating that wave phenomena played an inconsequential role in
the problem of motion in gravity, due to the small size of the effect
for known astronomical systems.

\section{The Bern and Chapel Hill conferences}

Between the war and the Bern conference of 1955 marking the 50th
anniversary of special relativity, general relativity was at a low ebb
(Eisenstaedt 1986a and 1986b). 
Work on the radiation problem seemed
confused and controversial, leading only to some
consensus that the problem required closer
attention. At the Bern conference Rosen, returning to the cylindrical
wave solution of his 1937 paper with Einstein,
adduced evidence backing up Scheidegger's position by proposing the
possibility that gravitational waves did not transport energy (Rosen 1955).
It is a peculiar characteristic of general relativity
that the energy contained in the
gravitational field, and thus the energy in gravitational radiation,
is not described in a coordinate invariant way. This energy is 
considered to be
real enough, and can be converted into other forms of energy which
can be expressed invariantly, but the principle of equivalence prevents
one from doing this for field energy in gravity. The reason is that any
observer in a gravitational field is always entitled to imagine himself
in a locally Lorentz (that is zero gravity) freely falling frame of reference
which, locally,  contains no field energy. Of course, one is not free to
transform away the entire field energy of a planet but one can
always choose co-ordinates on an infinitesimally 
small portion of its surface so as to
eliminate the field energy in that region. Thus it is said that gravitational
field energy is 
non-localizable.$^{16}$ This problem of defining field-energy had
led Einstein, Landau and Lifschitz and others to employ a non-invariant
quantity known as a pseudo-tensor to describe energy in the wave flux
in their back reaction calculations. Rosen now observed that each of
these (slightly different) definitions of the pseudo-tensor showed no
energy at all when applied to the cylindrical waves of his 1937 paper
with Einstein in cylindrical co-ordinates. 
Although drawing conclusions on the tentative basis
of the pseudo-tensor was 
regarded as dangerous, Rosen observed that the result
seemed to support the view of Infeld and Scheidegger. This
cast further doubt on the uncertain status of wave phenomena
in gravitation theory.$^{17}$ 

The Bern conference is remembered as an important stimulus to the
field of relativity. 
The discussions there, and the interest taken by Felix Pirani,
prompted Hermann Bondi to take up the problem of 
gravitational radiation.$^{18}$
Bondi
brought an open mind to the issue, in the sense that he was sceptical enough
of the existence of gravitational waves. He was influenced in this 
by Eddington, from whose writings he
learned relativity.
Eddington's emphasis on a coordinate invariant approach, making use of
tensorial quatities such as the Riemann curvature tensor, had enabled him
to show that certain classes of gravitational waves ``in existence'' before
1922 were spurious (Eddington 1922).
Bondi, like some relativists of the day, 
was not impressed by the existing radiation reaction work, finding
Landau and Lifschitz' treatment ``a little
glib''.$^{19}$ 
At the same time, gravitational waves seemed like an attractive
topic within gravitational theory, since in this area the
predictions of general relativity diverged radically from those of Newtonian
gravitational theory. Up to this time, most work in relativity, outside
of cosmology, had been devoted
to deriving small corrections to Newtonian theory, such as the famous
perihelion shift of Mercury, a more precise estimation of which was one of
the goals of the EIH paper (Robertson 1938). The study of gravitational waves,
if they existed, seemed likely to generate more ``new physics'' than simply
adding terms to Newton's theory.
 
Now, as Infeld himself observed when writing of his surprise at Einstein's
``proof'' that waves did not exist, no
respectable modern field theorist would, under normal circumstances, deny
the existence of radiation in a field theory. The mere fact that the
force was propagated in the field 
rather than by action-at-a-distance, a basic tenet of
all relativistic field theories, 
seemed to imply the existence of radiation. Einstein also
remarked, in his letter to Born, of the ``certainty'' which the analogy
between the linearized Einstein equations and electromagnetism had
inspired concerning the existence of a gravitational analogue to the
Maxwellian wave equation. Bondi nevertheless seized on a key argument
made by Infeld and Scheidegger, which seemed to him important.

As Scheidegger observed, relativity
occupied a ``peculiar place'' amongst 
classical field theories (Scheidegger 1953). 
One important peculiarity 
is that the equations of motion are constrained by the
field equations, as Einstein had noted. In electrodynamics,
where this was not the case, one was perfectly free to demonstrate damping
effects by moving the particles around in whatever fashion, and showing
that this gave rise, when the field equations were invoked, 
to radiation and loss of energy from the local system.
In relativity, it was necessary to show that the motions in question were
allowed by the same field equations. 
This was all the more important when one considered the question
of what {\it type} of motion gave rise to radiation. One obvious example
was an accelerating 
charge in electrodynamics. What of the apparently equivalent
case of a falling mass? It was clearly accelerating with respect to the
person who dropped it, but in a relativistic sense, it was merely following
a geodesic, doing what came naturally, as it were.$^{20}$ In terms of the local
spacetime, the particle that was really being {\it accelerated} was the
one still being held in the observer's other hand, which was prevented
from falling freely. Which one of these particles {\it ought} to radiate? This
question had no immediately obvious answer which the relativists of
the day could agree upon.$^{21}$ 

At the
Chapel Hill conference of 1957
and elsewhere at that time, Bondi pointed
out the distinction between two masses being waved about at the end of
someone's arms,$^{22}$ clearly not following geodesics, and clearly emitting
gravitational waves (but vanishingly weak ones!), and two masses
in a binary star system, following geodesics and, if Infeld and Scheidegger
were right, not radiating anything (De Witt 1957, pg. 33). 
Since gravitational forces were likely
to be the only forces capable of moving large masses very quickly, the
issue of whether purely gravitational systems could give rise to radiation
was an issue of whether such radiation would ever be detectable. That
issue, to the surprise of most theorists, was soon to become one of some 
practical interest.

The Chapel Hill conference
on ``The Role of Gravitation
in Physics'' brought together relativists 
and theoretical
physicists interested in then new topics such as quantum gravity.
The session on gravitational radiation was lively and varied.
Felix Pirani presented important new work on wave theory (De Witt 
1957, pg. 37). 
Influenced by the Irish relativist John Synge
during a year spent in Dublin (and also on the work of Petrov (1955)
and Lichnerowicz (1955)),$^{23}$ 
Pirani drew attention to the Riemann curvature
tensor, whose importance had previously been stressed by Eddington in
his 1922 paper,
as an invariant geometrical quantity which was well suited to the
description of the behavior of gravitational waves. Using 
the geodesic deviation description of gravitational effects
advocated by Synge, he showed how particles
in the path of a wave were moved about relative to each other 
by the spacetime curvature of the
passing wave. In this view, gravitational waves were depicted as ripples
in the fabric of spacetime itself, whose physical effects were observable
by monitoring the relative motion of two adjacent particles
during the passage of a wave.

Later in the conference 
an interesting exchange took place
during the section on quantization of gravity.
During Richard Feynman's 
presentation on the need for a quantum theory of gravity,
Rosenfeld made the following remark: 

\begin{quotation}
It seems to me that the question of the existence and absorption of waves 
is crucial for the question whether there is any meaning in quantizing
gravitation. In electrodynamics the whole idea of quantization comes
from the radiation field, and the only thing we know for sure how to
quantize is the pure radiation field. (De Witt 1957, p. 141)
\end{quotation}

Feynman demurred somewhat from
the premise, arguing that there existed a quantum theory of electrostatics,
but agreed that some of his arguments in favor of
quantization depended on the existence of waves. Bondi was moved to note
that ``this vexed question of the existence of gravitational waves does
become more important for this reason.'' Feynman then presented an argument
based on Pirani's earlier talk.
Appealing to 
the equation of geodesic deviation, he argued that a particle lying
beside a stick would be rubbed back and forth against the stick by a passing
wave, and the friction would generate heat, so that energy would have been
extracted from the wave. Furthermore, he felt that any system which could
be an absorber of waves, could also be an emitter. 
For these reasons, he expected gravitational
waves to exist (supplement to De Witt 1957).$^{24}$

This line of argument, suggested by Pirani's new work, was also
elaborated in two papers published that same year. 
In a letter to Nature, Bondi used a slightly different version of 
it to refute Rosen's argument of 1955 on energy transport
(Bondi 1957), as did Joseph Weber
and John Wheeler in a more detailed paper (Weber and Wheeler 1957).
Weber demonstrated real confidence in the physicality of gravitational waves
by embarking within a few years 
on an experimental program to detect them, using large resonant metal bars
as antennae (Weber 1960).
Quixotic is probably not quite the word contemporary theorists would
have used to describe Weber's aim.$^{25}$
The wave theory, in so far as it existed 
at all, with no particular notion 
as to potential astrophysical sources or signals, 
would be
better described as a ``disabling'' rather than an enabling theory
for experiment. The
quadrupole formula, the only guide to source strength and signal amplitude,
suggested that any waves reaching the detector would be very weak. 
With no theory of sources, the question
of what frequency to search at 
was theoretically undetermined.$^{26}$
It is remarkable that
the field of gravity wave detection began at a time when the
theoretical state of the subject was in such disarray.

\section{The Rebirth of Relativity}

An important requirement for the development of any scientific field is
funding. The field of gravitational wave theory was fortunate in this
regard in that, from 1956 to 1963, Joshua Goldberg was responsible for
United States Air Force support of research in general relativity, based
at the Aeronautical Research Lab at Wright-Patterson Air Force Base in
Ohio. At this time, and up until the passage 
by Congress in 1969 of the Mansfield Amendment
prohibiting 
the Department of Defense from sponsoring basic scientific research,
the US armed forces
provided considerable financial support for even 
very esoteric subjects in theoretical physics. Goldberg was active himself
in the study of gravitational radiation, as we have seen, and did much to
encourage groups such as that of Bondi and Pirani at King's College, London.
Although support was available for groups outside the US, it was not
permitted to support scientists based in communist countries, inhibiting 
the use of these funds to facilitate travel 
between the London group and Infeld's group in Warsaw, who interacted
extensively.$^{27}$ 
The Air Force laboratory itself was home to an active group until
the 1970s.
With one of his earliest grants, Goldberg was able to support
the Chapel Hill conference organized by Bryce De Witt 
with Air Force money, and this important
meeting became the forerunner of the successful 
General Relativity and Gravitation (GRG)
series of conferences, which continues today. For a valuable account
of this unlikely episode in the history of general relativity,
see Goldberg (1988). 

Following the Mansfield
Amendment, research in 
relativity theory in the US depended primarily for its support
on the National Science Foundation (NSF). From
1973 to the present, the chief advisor on funding for gravitation physics
at the NSF has been Richard Isaacson, like Goldberg a relativist who has
made important contributions to the theory of gravitational waves. Isaacson
had also previously worked at the Air Force 
laboratory on the Wright-Patterson base. By good
fortune then, despite the overall 
decrease in funding for theoretical physics
precipitated by the Mansfield Amendment, the principal source of funds
for research on gravitational wave theory remained in sympathetic and
knowledgeable hands.$^{28}$

As interest in relativity grew in the period after Chapel Hill, the
reaction problem was pursued with renewed vigor.
The EIH approximation was adopted by Andrzej
Trautman, a student in Infeld's
group in Warsaw, who departed from Infeld's approach in adopting
``outgoing wave only''
boundary conditions. He also confirmed Goldberg's earlier claim that the net
back-reaction effect could not be transformed away, merely moved from one
point in the expansion to another. 
He found positive damping, although differing
somewhat from the quadrupole formula result (Trautman 1958a, 1958b). 
Infeld himself stuck to his earlier opinion, despite the contrary
views of his students. In his 1960 book, {\it Motion and Relativity}, he
included a detailed argument against the existence of back reaction
in freely falling systems (Infeld and Plebanski 1960), 
without the knowledge or agreement of
his co-author and former student, Jerzy Plebanski.$^{29}$ 
Another effort at this
time,
by Peres, initially found anti-damping, as had Hu, 
but this was corrected
shortly after, and his new result agreed with that of Landau and 
Lifschitz for circular binary orbits (Peres 1959,1960). Peres'
second paper has been 
been referred to as containing 
the first correct back reaction calculation 
(Thorne 1989). Nevertheless, the perceived
arbitrariness of the slow-motion approach in imposing the wave
zone boundary conditions from one step in the expansion to the 
next, which seemed reflected in the wildly differing
results produced by the method, gave rise to arguments that the approach was
hopeless (Bonnor 1963).

While conceptually more appealing in some ways, 
the alternative fast-motion approach,
as developed by Havas and Goldberg (Havas and Goldberg 1962) and others
(for example, Bertotti and Plebanski 1960, Kerr (1959), Westpfahl
(1985))
was also proving 
frustrating. It was a difficult task to go beyond the leading
order corrections to the linearized theory 
and the results of applying that step to the reaction 
problem, published by Smith and Havas,
again showed an energy gain in the source (Smith and Havas 1965). 
Therefore many theorists at the time concluded
that the question of whether
freely falling sources experienced damping remained unsettled.

Bondi, who with his collaborators had done much to improve the understanding
of wave propagation far from the source (see especially Bondi, van
der Burg and Metzner 1962 and Sachs 1962) 
made this point at the Warsaw conference of 1962 (Bondi 1962).$^{30}$ 
However, there were those, like Feynman, who viewed
the relativists' caution 
with impatience. Feynman was ``surprised to find a whole
day at the conference devoted to this question'' (of whether gravity
waves could carry energy), as far back as Chapel
Hill (letter from R.P. Feynman to Victor Weisskopf, February 11, 1961)$^{31}$,
and was caustic in his appraisal of the discussions
at the Warsaw conference, noting they were ``not good for my blood
pressure'' in a letter to his wife (Feynman 1988).
Bondi's lecture, however, inspired the astrophysicist 
Subrahmanyan Chandrasekhar
to take up the problem.$^{32}$
Throughout the 1960s, Chandrasekhar developed his own slow-motion
formalism, dealing with extended fluid bodies (as opposed to point
masses) at one
post-Newtonian order after another 
(Chandrasekhar 1965).
By the end of the decade he had advanced far enough in the expansion (to
order $(v/c)^5$ beyond Newtonian order)
to describe reaction effects. His conclusion agreed with the
quadrupole formula result (Chandrasekhar and Esposito 1970). At about this time
William Burke, a student of Kip Thorne's at Caltech, 
introduced improvements to the slow-motion approach which 
removed much of the arbitrariness in imposing the boundary conditions.
Burke made use of the applied mathematics technique of matched
asymptotic expansions, which allowed one to determine the solution
to the problem of motion in the zone near the source, by matching it
through an intermediate zone, to the ``outgoing wave only'', or other
potential of choice, in the far zone of the waves. 
In this way the chosen boundary condition
could be unambiguously applied to the solution of the near zone problem,
thus addressing the arbitrariness which bedeviled the slow
motion approach up to this time (Burke 1969). 
Employing Burke's novel approach, Burke and
Thorne also derived the quadrupole formula for emission from binary 
systems (Burke and Thorne 1970).

During the sixties, great progress had been made on many fronts
in the description of wave propagation and interaction with matter. 
Possible astrophysical sources, such as supernovae and binary
neutron stars, began to be suggested, 
inspired at first by Weber's work (Dyson 1963).$^{33}$ 
Some
experts were of the opinion that the subject was maturing and furthermore
the prospect of some real astrophysical application
for gravity waves, seemed to emerge with the discovery of the
quasi-stellar (``quasar'') radio sources (Fowler 1964; Robinson, Schild
and Schucking 1965 and Cooperstock 1967). 
Then, to the great surprise of the
theoreticians, Weber announced in 1969 that he was detecting gravitational
waves (Weber 1969). 
Although his results, which confounded all theoretical predictions
of source strengths then and since, were eventually discounted amidst
much controversy, 
they focused much attention on
the subject, and sparked a great increase in the number of experimentalists
working on gravitational waves. 
(See Collins 1975 and 1981 for a detailed account, and 
Franklin 1994 for an alternative viewpoint). 
On the theoretical front,
research
in the 1960s on black holes, cosmology and other topics had
made the field of relativity very relevant to astrophysics. Gravitational waves shared
somewhat in this popularity, 
and seemed likely to continue to grow
in practical importance as experimental interest waxed. The discovery
of the first binary pulsar (PSR 1913+16) by Hulse and Taylor in 1975
(Hulse and Taylor 1975)
crystalized the excitement in the field, providing the first test
bed for strong field effects of general relativity, although there
were doubts at first that the system would exhibit measurable
orbital damping effects (Damour and Ruffini 1974).$^{34}$

\section{The Quadrupole Formula Controversy}

The successes in improving the slow motion approximation, and the
increasing likelihood of practical applications of gravitational radiation
theory encouraged some experts, such as Kip Thorne, to suggest that
the reaction problem was now well understood,$^{35}$ 
and the multipole
formalism could be used with confidence in astrophysical applications to give
approximate estimates of source strength, much as one would 
in electromagnetic wave theory (Thorne 1980). 
This viewpoint however,
was sharply opposed by some others who were still seriously dissatisfied
with the state of the field.$^{36}$
They were particularly concerned that the 
quadrupole formula
would be used as a reliable formula in contexts in which its results
might be wholly misleading. One of these
was Havas who was still very unhappy with the various
slow motion results (Havas 1973).$^{37}$
One of his students, Arnold Rosenblum, brought the unsatisfactory
state of affairs to the attention of the mathematical
physicist J\"{u}rgen Ehlers, who also took up the cause of alerting the
relativity community to the dangers of complacency on the matter.$^{38}$

The alarm sounded by Havas, Rosenblum and Ehlers had the effect
of again focusing attention on the reaction question, and this
interest was redoubled by the announcement, in 1980 of observations
of orbital decay in the binary pulsar. Taylor and
coworkers, after years of careful observation of the system, were
able to announce an orbital period decrease in line with the
predictions of the quadrupole formula with an accuracy of measurement
of about 20\% (Taylor and McCulloch 1980). 
The warnings of the unverifiability of the quadrupole
formula within the
theory now had their effect. A chief use of the binary pulsar data since its
discovery had been as a test of general relativity
 against rival theories of gravity.
Agreement between observation and the quadrupole formula 
could only constitute
a test of general relativity theory if the quadrupole formula 
was established as a prediction of the theory.$^{39}$
Not all relativists were of the opinion that it was 
so established.$^{40}$ There
was a surge in interest in the problem of motion and in back-reaction
in particular, including by some who had not previously
worked on the problem. The great majority of the new results vindicated
the use of the formula. A further round of sharp debate ensued as
these results, via a wider variety of approaches then ever
before, and in greater detail than ever before, convinced many that
the issue was at last settled,
pushing the remaining sceptics into
an embattled minority. As the eighties advanced, and the increasingly
convincing experimental data continued to agree solidly with the
theoretical work carried out in close parallel by Thibault Damour
and his collaborators (Damour 1983), the debate slowly died away. 
At present the focus in the field is on calculating higher order
contributions to the waveforms produced by binary systems for use in     
conjunction with data extraction techniques in the next generation
of wave detectors, indicating a high degree of confidence in most
quarters in the basic theoretical position. A direct detection of
gravitational waves is still being sought, and this is one of the
principle goals of the planned new detectors.

An important feature of the radiation reaction debate in the seventies
and eighties was the series
of review papers by different authors, each employing 
the history of the subject 
to illustrate a particular view of the contemporary state of the field.
These papers show that relativists were keenly aware of the history
of their field and
they were able to draw lessons from their reading of history which
reinforced the points they wished to make. The earliest of these
papers was that of Ehlers, Rosenblum, Goldberg and Havas whose
argument was that previous attempts to deal with the back-reaction problem
were all inadequate in one way or another. In consequence, they
advanced an
outline of a program which would overcome these past failings
(Ehlers, Rosenblum, Goldberg and Havas, 1976). 
Essentially an attempt to formulate a research program
for the subject, their paper was followed by an Enrico Fermi
summer school in Varenna organized by Ehlers, 
whose aim was also to foster new work in the
field along more rigorous lines than before (Ehlers 1979).

Walker and Will in 1980 took a very different tack, addressing the
problem of non-reproducibility which had plagued the subject (Walker and Will
1980).
They argued that a basic iterative 
algorithm, applicable for both fast motion and
slow motion methods, could be followed to recover the quadrupole
formula from reaction
calculations. They presented an analysis of a cross section of well-known
calculations, dating back to the paper of 
Hu in 1947, and argued that those which had advanced
through sufficient steps in the iteration recovered the quadrupole
formula, and that others,
with fewer steps did not (except for a couple which found the result
with
the aid of compensating errors). In this view of the history
of the field, there existed
a definitive method by which the standard results could be recovered
in a reliable way. This was in stark contrast to the views expressed by
Ehlers et al., which were
to advocate a more general prescription, whose outcome 
was not yet known. Yet another view was put forward by Cooperstock and
Hobill in 1982. They refused to set forward a general scheme or advocate
a particular result, instead arguing against preconceived notions
(Cooperstock and Hobill 1982). Their
history, as befitted their standpoint, was more descriptive than prescriptive,
celebrating the diversity in the development of the field. 
Another protagonist with an interest in and excellent knowledge of the
field's history was Damour. His papers were often prefaced with a discussion
setting his work in a historical context
(for example Damour 1982). In this role, the object of
history was to motivate the new work being presented, and the focus was
on the previous failings which were being addressed by the new contributions
(see, for instance, Damour 1983).
A more active role for the historical literature was found in the account
of James Anderson, who returned to the Einstein-Infeld-Hoffmann scheme complete
with its surface integral method, and married it to the matched  
asymptotic expansions of Burke, with further additions of his own,
to produce another influential 
derivation of the quadrupole formula (Anderson 1987).

A very significant aspect of the debate in the seventies and eighties
was the problem of when theory ends.$^{41}$ As we have seen, different authors
could look at the same history and give very different answers to this
question. One answer might be, it already has ended, we really know the answer
(``Conservative'').
Another is, it has just ended now, with this paper, for the issues
addressed (``Technocratic''). 
A third is, it will end, as soon as the general program we
advance is carried through (``Marxist''). 
A fourth is that it can never end, and it
is best that it should not (``Anarchist''). Finally there is the view that
the answer is hidden in the past, waiting to be extracted and pieced together
from the literature (``Archaeological''). It is interesting that just as
there was agreement on the details of the history (and the debate was
largely a historical debate), opinions diverged on the matter of
{\it interpretation}. The lesson of history was different for everyone.
This is still the case, but the debate having lost its impetus, the
individual perception of history has lost its public relevance once more.
The dynamic of the debate is that some level of consensus must be found for 
the resolution of an existing problem,
and yet progress seems to be measured by many scientists
by the extent to which an issue can be settled, allowing the next problem
to be addressed. A field like General Relativity has historical memories of the 
isolation which may be the fate of a discipline which does not progress
in this way. The remarks of Feynman at Chapel Hill (De Witt 1957, pg. 150),
express the view of the progressives, when he says ``the second choice of
action is to ... drive on,'' to 
``make up your mind [whether gravitational radiation exists] and 
calculate without rigor in an exploratory way''. He concludes with 
the advice, ``don't
be so rigorous or you will not succeed.''$^{42}$
The contrast in attitude suggested here 
may explain why the debate tended to become
more vitriolic in its last stages, as a consensus developed for many, with
some still arguing that the matter was unsettled.$^{43}$ 

In studying the controversy following Weber's announcement of gravitational
wave detections, Harry Collins (1985) has introduced the concept of the
Experimenter's Regress. This describes the difficulty faced by experimenters
when confronted with a dispute over non-confirmation of claimed results. Since
none of the experiments will exactly duplicate the others' behavior, achieving
consensus is hampered by the problem that the device which is working properly
should get the correct result, but the correct result can only be known from the
output of a properly operating device. Although Collins' view has been
criticized (Franklin 1994), it seems to provide 
a useful model for understanding the Weber controversy. In the theoretical
controversy surrounding gravitational waves, one seems to observe a similar
phenomenon,
the ``Theoretician's Regress''. The complex, tedious calculations designed
to approximate to the full general relativity theory can be thought of as
experiments, with the theory itself in the role of a notional ``reality''.
These experiments constituted a delicate technical apparatus, designed to
probe this ``reality'', aided by the craft and mathematical skill of the
theorists. ``Experimental error'' was impossible to account for fully, whether
as systematic error in the form of an inappropriate expansion scheme or failure
to properly control errors from neglected terms (a difficult problem which 
was rarely addressed 
programmatically), or as accidental error in the form of simple calculational 
mistakes amidst the welter of terms which had to be collected. 

As with the
experimentalists, direct replication of another method was 
rarely even attempted. Even the best known schemes, such as EIH, were employed
with improvements designed to simplify the calculations
or overcome objections in principle, 
such as the use of point mass sources (Anderson 1995).
Therefore, the array of review papers, conference workshops and other social
efforts to achieve consensus had to overcome the cycle of regression 
constructed by the fact that the right scheme would be the one which gave the
right result, but the right result was the answer given by the right scheme.
The difference in emphasis between those who gave weight to having the right
answer, and those who preferred to rely on method alone gave rise to further
disagreement. One event which helped to partially break the cycle was the
advent of the binary pulsar data. Initially this gave rise to more activity
and more disagreement, but it also lent external support to the preferred
``right result'' given by the quadrupole formula. It did not however, put an
end to disagreements about the correctness of various methods, except in so
far as it tended to rule out methods which disagreed with the canonical result.
This was enough to gradually bring an to end the public side
of the quadrupole formula controversy.

Throughout all this, one notes the 
tensions within the field over technical matters,
especially regarding the level of rigor required
to inspire confidence in a particular result. Relativity has a tradition
which places it towards the mathematical end of the spectrum in this
regard amongst branches of theoretical physics. Yet from the sixties
on, astrophysics and relativity became relevant to each other, even
spawning the new field of relativistic astrophysics. Theoretical
astrophysics stands at the opposite extreme from relativity, preferring
a more ``physical'' approach, eschewing not only mathematical
rigor, but also dependence on exact results. Order of magnitude
calculations and heuristic arguments 
are common. Such arguments, for instance, might be used to identify the
``correct'' result, as a guide when undertaking longer calculations.$^{44}$ 
Within relativity there were those whose
practice tended towards each approach, and it was naturally difficult
for them to agree on the question of standards of proof.$^{45}$ 
For practical purposes results such as the binary
pulsar 
measurements were obviously welcome, but at issue was on whose
terms a given result was to be accounted a prediction of general
relativity: the
``astrophysicists''
or the ``mathematicians''.

A second, less significant, mixing of fields
concerned attempts to quantize gravity.
Especially in the fifties, it was argued by some that the 
existence of radiation
was a crucial matter for this project (Rosenfeld in De Witt
1957, pg. 141; Rosen 1979). In fact, Felix Pirani's view was
that ``the primary motivation for the study of [gravitational radiation]
theory is to prepare for quantization of the gravitational field.'' 
(Trautman, Pirani and Bondi 1965, pg. 368).
The uncertain position of
general relativity as an independent, yet thriving field, seems to
have played into fears and attitudes concerning the radiation problem.
Relativists' own practices and their own opinions of what the problems were
in the field may have seemed endangered by the twin possibilities
of classical relativity becoming a mere adjunct to astrophysics,
and the theory as a whole being submerged by a unified
quantum field theory of gravity (Roger Penrose in Lightman
and Brawer 1990, p. 429).
The emergence of relativity into the mainstream of physics had a
highly ambivalent aspect for relativists, in that it brought with it the danger
that the character of the small stream would be lost in the larger
current. The fears and hopes which this dual prospect raised for
scientists who had consciously chosen the field for its own beauty 
and intimacy
no doubt helped shape attitudes in the debate.

\section{A Final Note}

I have concentrated, in this paper, on the emergence of certain
important issues which contributed to the uncertainty and controversy which
at times surrounded the theoretical development of the subject of
gravitational radiation. In doing so, not only have I focused here
on the immediate post-war period up to about 1960,
but I have deliberately not attempted to cover the entire breadth
of the literature for any period. 
I have tried to 
illustrate how the debate on the existence of gravitational waves came
to arise as a serious discussion, and how the subject's own history was
used as a rhetorical and motivational tool in subsequent debates after
its emergence as the subject of experimental and not just theoretical
research. I have left many interesting aspects of the history of this
problem for another time.
For now,
I have tried to give a sense of how the thought that gravitational
waves might NOT exist first arose, and then became a serious issue in the
field of relativity, and how the issues raised fed into later debate as the
subject matured. It is worth noting that {\it scepticism} about their existence
encouraged important scientists, such as Bondi, to focus attention on 
gravitational waves, at a time when those who were certain of their existence
dismissed their effects as insignificant (Landau and Lifschitz and Fock, 
for instance). If
the attempt to detect gravitational radiation is now a multi-million dollar
field, thanks to the pioneering work of Weber on the experimental side, some
credit must go to those who thought the theory of the subject worth advancing
for reasons of principle many decades ago.

This paper makes considerable use of interviews with participants in
the events discussed. As yet, no arrangements have been made for a
permanent disposal of the materials from these interviews, and the tapes
have not yet been made into complete transcripts. In the meantime, 
anyone interested in their contents should contact the author directly,
who will be happy to oblige any requests, subject to the agreement of the
interviewee. In one or two cases, such as the interview with S.
Chandrasekhar, no tape is available, only notes taken at the interview.

\section{Acknowledgements}

My most grateful thanks go to Diana Barkan, for her help and support
with this research, and to Kip
Thorne, at whose suggestion I undertook it. 
Thanks also peter Havas, John Stachel, Jean Eisenstaedt and Martin
Krieger for valuable discussions.
I am also grateful for permission granted 
by the Albert Einstein Archives, The Hebrew University of Jerusalem, 
as well as by the Einstein 
papers project and John Tate jr.,
to quote from the Einstein-Tate correspondence, and to
Robert Schulman for his kind help at the Einstein papers project in Boston.
I am indebted to the Archives at the California Institute of Technology
for permission to quote from Robertson's letter to Tate.
To the organizers of the Berlin conference, for their hospitality, kindness
and generosity, especially to Tilman Sauer and J\"{u}rgen Renn, my thanks.
I am also the grateful recipient of a Doctoral Dissertation Improvement
grant (No. SBR-9412026) 
from the National Science Foundation, which enabled me to travel to
consult archival material and conduct interviews. Finally,
I thank all of those who agreed to be interviewed for their time and
help.

\section{Notes}
$^{1}$ The construction of history as part of the self-definition of a
field of science is an important topic in the history of science. For an
excellent discussion in a different context, see Barkan (1992).

$^2$ The original reads
\begin{quotation}
Ich habe zusammen mit einem jungen Mitarbeiter das interessante
Ergebnis gefunden, da{\ss} es keine Gravitationswellen gibt, trotzdem
man dies gem\"{a}{\ss} der ersten Approximation f\"{u}r sicher hielt.
Dies zeigt, da{\ss} die nichtlinearen allgemeinen relativistischen
Feldgleichungen mehr aussagen, bezw. einschr\"{a}nken, als man bisher
glaubte (Born, 1969).
\end{quotation}
The translation is by Irene Born, from the English language edition.

$^3$ Although the original version of Einstein and Rosen's paper probably
no longer exists, its original
title is referred to in the report by the {\it Review}'s referee
(EA 19-090).

$^4$ The translation from the original German is by Diana Barkan. The
emphasis in the letter is Einstein's.

\begin{quotation}
Sehr geehrter Herr:

                    Wir (Herr Rosen und ich) hatten Ihnen unser
Manuskript zur \underline{Publikation} gesandt und Sie nicht
autorisiert, dasselbe Fachleuten zu zeigen, bevor es gedruckt
ist. Auf die - \"{u}brigens irrt\"{u}mlichen - Ausf\"{u}hrungen
Ihres anonymen Gew\"{a}hrsmannes einzugehen sehe ich keine
Veranlassung. Auf Grund des Vorkommnisses ziehe ich es vor,
die Arbeit anderweitig zu publizieren.

                               Mit vorz\"{u}glicher Hochachtung

P.S. Herr Rosen, der nach Sowjet-Russland abgereist ist, hat
mich autorisiert, ihn in dieser Sache zu vertreten.
\end{quotation}

$^{5}$ In a letter to Einstein in March 1936, Cornelius Lanczos remarks
on ``the rigorous criticism common for American journals'', such as the
{\it Physical Review} (translated and quoted in Havas 1993, pg. 112). 
Infeld claims
that the German attitude, by contrast, was ``better a wrong paper than
no paper at all.'' (Infeld 1941, pg. 190). 
Jungnickel and McCormmach (1986) describe the editorial       
workings of the {\it Annalen der Physik} in the first decade of
this century in some detail. They note that 
``the rejection rate of the journal was remarkably low, no higher than
five or ten percent'', and describe the editors' reluctance to reject
papers from established physicists (pg. 310). As this was the time and
place in which Einstein began his published career, the ``rigorous criticism''
he was to experience very 
shortly after receiving Lanczos' letter must have come as
something of a shock.

$^{6}$ Einstein's bibliography to 1949, given in Schilpp (1949) lists no
papers by him appearing in the {\it Review}
after 1936, and the index of the {\it Physical Review}
from then until his death refers only to one short note of rebuttal, mentioned
by Pais (1982) in his 
brief account of the rejection of the Einstein-Rosen paper.

$^7$ The paper appeared in the Franklin Journal under a different
title and with radically altered conclusions in early 1937. That it had
previously been accepted in its original form is indicated by a letter
from Einstein to its editor on 13/11/36 (EA 20-217), 
explaining why ``fundamental''
changes in the paper were required because the ``consequences'' of the
equations derived in the paper had previously been incorrectly inferred.

$^{8}$ Curiously, Infeld states that when he communicated to Einstein
his discovery with Robertson of an error in his (Infeld's) version of the 
proof, Einstein
replied that he had coincidentally and independently uncovered a (more
subtle) error  
in his own proof the night before (Infeld 1941, pg. 245). 
He does tell us that Einstein's
position still had to evolve from that of demolishing his proof, to that
of reversing it (by 
showing an exact solution for cylindrical waves), and this was
Robertson's key contribution according to 
Rosen's paper of 1955.
Unfortunately, Infeld
gives us no details of the false proofs and their correction in his account, 
which was intended for a popular audience. He does relate the amusing
detail that Einstein was due to give a lecture in Princeton on his new
``result'', just one day after completely reversing his conclusions on
its validity. He was forced to lecture on the invalidity of his proof,
concluding by stating that he did not know 
whether gravitational waves existed or not
(Infeld 1941, pg. 246).

$^9$ The identity of the Review's referee
is unfortunately not known. Few records of the journal exist for this
period, and the report has only survived amongst Einstein's own papers. 
It is 10 pages long and shows an excellent, if not perfect, familiarity with
the literature on gravitational waves (the referee knew of
Baldwin and Jeffrey's 1926 paper, but not Beck's of 1925). 
The copy forwarded to Einstein
is typewritten and the spelling follows American practice (``behavior''
rather than ``behaviour'', ``neighborhood'' rather than ``neighbourhood''). 
It is likely, therefore, that the author was an American with a strong
interest in general relativity, not a very inclusive category at this time.
It is tempting to suspect Robertson himself, but there is nothing to support 
this in his surviving (and extensive) correspondence with Tate, apart
from the one letter quoted below, whose evidence is persuasive but not
conclusive.

$^{10}$ Interviews by the author with Hermann Bondi (November 7, 1994) 
and Felix Pirani (October 25, 1994). 
Pirani reviewed the McVittie (1955) paper for
{\it Mathematical Reviews} and was dissatisfied with its conclusions
(Pirani, 1955).

$^{11}$ In their work, Bondi, Pirani and Robinson followed the new
approach of Lichnerowicz in imposing regularity conditions on the
metric (Lichnerowicz 1955). For a thorough 
review of the tangled
history of plane gravitational waves, see Schwimming (1980).

$^{12}$ See also 
Damour (1982) for a brief but interesting discussion of Laplace's
``radiation reaction'' calculation. It is now known, from laser
range finding, that the moon is receding from the earth, not approaching
it. But the increased lunar orbital angular momentum is gained at
the expense of earth's rotational velocity, by tidal friction. 
The resultant lengthening of the earth's day gives the appearance
of quickening to all celestial motions, including the lunar orbital 
period (i.e. although the month has lengthened, it is shorter in
terms of days, since the day has also grown longer).

$^{13}$ Since 
general relativity is a non-linear theory,
the fact that two potentials 
(the advance and retarded) satisfy the field equations
does not imply that their linear 
combination (half advanced plus half retarded)
would, as it does in electromagnetism. 
In linearized gravity, however, this obviously does follow.

$^{14}$ See Havas (1989) for an excellent review. 

$^{15}$ Interviews with Joshua Goldberg (April 10, 1995) 
and Peter Havas (April 5, 1995).

$^{16}$ In 1968, Richard Isaacson 
discovered an invariant tensorial quantity which
described wave energy in a 
local sense, by averaging over a wavelength of the wave.
Thus, using this approach, 
gravitational wave energy can be localized within a
wavelength, but no further (Isaacson 1968).

$^{17}$ As we shall see, Rosen's paper was soon answered in a manner
convincing to most relativists. He himself revised his opinion on this matter
in a letter to the Physical Review (Rosen 1958), after realising that,
when using Cartesian co-ordinates, the pseudo-tensor did show energy in
the cylindrical waves. His new calculations
on the energy content 
of cylindrical waves did not appear until after some delay (Rosen
and Virbhadra 1993). The issue was addressed in some depth in
the fifties, however (Stachel 1959).
The problem of the pseudo-tensor in the
study of gravitational waves was not new then, nor has it entirely ceased to be
the subject of debate since. In recent years, Fred Cooperstock has suggested
that, based on the hypothesis that the preferred frames of reference 
when describing the field energy should
be those which eliminate the pseudo-tensor,
the gravitational field energy should be described only by an invariant tensor
quantity. The result of this would be that the conservation relation in 
relativity would require that no field energy be present where there was no
matter, preventing gravitational waves from propagating energy through empty
space (Cooperstock 1992). In the very early days of general relativity 
Levi-Civita made a proposal with somewhat similar (but more drastic) 
consequences,
in response to the confusing and
incorrect results derived by Einstein in his 1916 paper on gravitational waves
(Levi-Civita 1917). For a very interesting discussion of this episode, which
includes some revealing comments reflecting the initial unease about
gravitational radiation brought on by Einstein's early errors
(including the mistaken conclusion of his 1916 paper that spherically
symmetric motions of matter could generate gravitational waves), see Cattani
and De Maria (1993).

$^{18}$ Interviews by the author with Felix Pirani (October 25, 1994) 
and Hermann Bondi (November 7, 1994).

$^{19}$ Interview with Hermann Bondi (November 7, 1994).

$^{20}$ The question of whether particles following geodesics should radiate,
given that they are behaving ``naturally'' in a gravitational field, seems
intriguingly Aristotelian.

$^{21}$ Interview with E.T. Newman (April 11, 1995). 
He relates how J.A. Wheeler
once asked a roomful of relativists to vote on the answer to the
two particle question and recalls the room being fairly equally
divided. This seems to be a rare example of the ``Democratic'' approach
to science.

$^{22}$ A number of those interviewed 
by the author recalled Bondi vigorously demonstrating
this method of generating gravitational waves.

$^{23}$ Interview with Felix Pirani (October 25, 1994).

$^{24}$ Cooperstock's 1992 paper (see note 10) 
contains an argument based on a counter-example to the Feynman-Bondi
thought experiments, which claims that no energy is deposited in the
``absorber'' despite the motion locally induced by the wave. 
His hypothesis would imply that waves 
exist in general relativity, are detectable
by certain types of instruments, but carry no energy.
However, this
paper and its conclusions have provoked little debate. 
This perhaps reflects the difficulty in physics of reopening an argument 
considered closed by most in the field. At some point, the premise of the
paper 
becomes sufficient grounds for dismissal. However, the problem may
be simply due to the fact that papers outside the current thrust of research
interests are unlikely to receive much attention, whatever their conclusions.

$^{25}$ Several interviews (especially one with Joseph Weber June 20, 1995)
and anecdotal recollections, 
as well as the impression given by conference proceedings,
agree that the reaction to Weber's initial efforts to detect gravitational waves
in the 1960s ranged between polite scepticism and derision.

$^{26}$ Interview with Joseph Weber (June 20, 1995).

$^{27}$ Interview with Felix Pirani (October 25, 1995).

$^{28}$ The advantage of having an insider at the primary funding agency
did not ensure that everyone in the field was sponsored to the extent
that they desired or felt necessary. 
Complaints about the funding choices made and
its effect on research directions were very noticeable
on the experimental side, where groups and research programs
depended very heavily on the munificence of different (usually
governmental) funding agencies. But even on the theoretical side, work on
the problem of motion or radiation reaction was computationally so intensive
that funding for postdocs and assistants could make a big difference
to a group or research program. It may be that less popular research programs
suffered in this regard (such as fast motion approximations versus slow
motion ones), but it is difficult to assess the extent of this factor. 
This partial assessment is based on interviews by the author with
Richard Isaacson (April 7, 1995), Joshua Goldberg (April 10, 1995), 
Peter Havas (April 5, 1995) and Joseph Weber (June 20, 1995). 
Given the importance of
debates during conference sessions, it is also worth noting the
complaint that,
because of the influence of slow motion advocates such as Infeld on the
organizing committee, the fast motion approximation was not discussed at
any of the GRG conferences, such as Warsaw 1962 (Peter Havas, private
communication). Thus, the growth of this research program may have been
retarded by a lack of exposure.

$^{29}$ Interview with Jerzy Plebanski (June 30, 1995). 
In general, however, Infeld proved reasonably
tolerant of the opposing viewpoints within his group. Indeed, in the
late sixties, shortly before his death, he was finally 
won over by his students'
arguments (interview with Andrzej Trautman October 17, 1994).

$^{30}$ Bondi, van der Burg and Metzner (1962) and Sachs (1962) showed
that when a certain function (known as the Bondi news function) was
present, an isolated system would lose mass to the emission of gravitational
waves. At Warsaw in 1962, in the
discussion with 
Bergmann and Feynman (Bondi 1962), Bondi stresses the importance
of dealing with specific equations of state in the components of the
binary system, because it was his opinion that binaries composed
entirely of pressure free dust would not radiate, as all particles
would follow geodesics and there would be no possibility of
``news'', in the form of a
departure from geodesic motion. In the case of a real physical system, even
if no deviation from geodesic motion occurs, this is news,
since no news, if not good news, is still news, if news was expected.
Bondi eventually decided against
his position that idealized dust filled binaries might not radiate
(interview with Bondi November 7, 1995). 
In the same discussion in the Warsaw proceedings
(Bondi 1962) Feynman gives a brief account of his own unpublished 
calculations which
convinced him that gravity waves exist.

$^{31}$ A copy of this letter was kindly supplied to the author by Kip
Thorne. Copies are also kept amongst the Feynman papers at Caltech.

$^{32}$ Interview with Subrahmanyan Chandrasekhar (July 12, 1995).

$^{33}$ Interview with Joseph Weber (June 20, 1995). 
Weber recalls that Freeman Dyson
suggested asymmetric collapse of stars during supernova events as one possible
source for his detectors in the early 1960s.

$^{34}$ Interview with Thibault Damour (October 11, 1994).

$^{35}$ Interview with Kip Thorne (July 17, 1995). 
Thorne recalls first putting this
view forward at a meeting in Paris, June, 1967.

$^{36}$ Interview with Kip Thorne (July 17, 1995). 
He recalls Havas taking issue with
his comments at the Paris meeting, June 1967.

$^{37}$ The non-linearities still continued to bedevil
the problem in some people's minds. 
At Caltech, despite Thorne's complacency,
Burke noted in early versions of his work that his
approach was not guaranteed to work outside of linearizable systems, and
therefore could not settle the issue for freely gravitating systems. There
is still on display at Caltech the record of a wager between Burke and Thorne
on whether non-linear effects would ``significantly affect the radiation
in the lowest order'' from sources in free-fall motion. Thorne gave odds
of 25-1 for this bet, which Burke conceded in 1970.

$^{38}$ Interview with J\"{u}rgen Ehlers (October 14, 1995).

$^{39}$ The rate of energy emission predicted by the quadrupole formula
can be writen as $dE/dt=(1/5) \langle (d^3 {\cal I}_{jk}/dt^3)^2 \rangle$,
where ${\cal I}_{jk}$ is the Newtonian mass quadrupole moment of
the source (evaluated at retarded time). The square of the third
time derivative of this quantity, averaged over several wavelengths
(the meaning of the angular brackets),
determines the total energy flux in the waves from the source.

$^{40}$ Interview with James Anderson (April 3, 1995).

$^{41}$ The analogy to the problem of {\it How Experiments End} 
(Galison 1987) should
be obvious. 

$^{42}$ The alert reader will have guessed that I have just described as
``progressives'' the same class of people whose historical outlook I
earlier labelled ``conservative''.
In this case, conserving and defending the orthodox historical account plays
a crucial role in the progressive agenda, discouraging debate on topics
which are regarded as settled and directing energy towards problem solving     
work within the established paradigm. 

$^{43}$ Interview with Fred Cooperstock (June 26, 1995). 
A number of other interviewees recalled
rather heated exchanges taking place at conferences in the early 1980s 
during the quadrupole formula controversy.

$^{44}$ This question of ``style'' in physics seems an important one. 
Chandrasekhar suggests that the greatest physicists (such as Newton)
employed both of these styles
equally well. He relates that Fermi would say that he would not believe
a physical argument without a mathematical derivation, nor would he believe
the mathematics without a physical explanation. Interview with S. 
Chandrasekhar (July 12, 1995).

$^{45}$ Interviews with Kip Thorne (June 14, 1995) 
and J\"{u}rgen Ehlers (October 14, 1995).

\section{References}

Anderson, James L. (1987). ``Gravitational radiation damping in systems
with compact components'' {\it Physical Review D} {\bf 36}, 2301-2313.

Anderson, James L. (1995). ``Conditions of Motion for Radiating Charged
Particles'' (preprint, Stevens Institute of Technology, Hoboken, New Jersey).

Baldwin, O.R. and Jeffery, G.B. (1926). ``The Relativity Theory of Plane
Waves'' {\it Proceedings of the Royal Society of London, series A}
{\bf 111}, 95-104.

Barkan, Diana Kormos (1992). ``A Usable Past: Creating Disciplinary Space
for Physical Chemistry'' in {\it The Invention of Physical Science} eds.
Mary Jo Nye, Joan L. Richards and Roger H. Stuewer (Dordrecht, Boston)
pg. 175-202.

Beck, Guido (1925). ``Zur Theorie bin\"{a}rer Gravitationsfelder''
{\it Zeitschrift f\"{u}r Physik} {\bf 33}, 713-728.

Bertotti, Bruno and Plebanski, Jerzy (1960). ``Theory of Gravitational
Perturbations in the Fast Motion Approximation'' {\it Annals of Physics}
{\bf 11}, 169-200.

Bondi, Hermann (1957). ``Plane Gravitational Waves in General Relativity'' 
{\it Nature}, {\bf 179}, 1072-1073.

Bondi, Hermann (1964). ``Radiation from an isolated system'' in
{\it Relativistic Theories of Gravity}, proceedings of Warsaw conference, 
July 25-31, 1962
ed. Leopold Infeld (Gauthier-Villiers, Paris) pg. 120-121. 

Bondi, Hermann, Pirani, Felix A.E. and Robinson, Ivor (1959). ``Gravitational
waves in general relativity III. Exact Plane Waves'' {\it Proceedings of
the Royal Society of London, series A} {\bf 251}, 519-533.

Bondi, Hermann, van der Burg, M.G.J. and Metzner, A.W.K. (1962). 
``Gravitational Waves in General Relativity VII: Waves from axi-symmetric
isolated systems'' {\it Proceedings of the Royal Society of London, series
A} {\bf 269}, 21-52.

Bonnor, W.B. (1963). ``Gravitational Waves''
{\it British Journal of Applied Physics}, {\bf 14}, 555-562.

Born, Max (1969). {\it Briefwechsel 1916-1955} (Nymphenburger
Verlagshandlung, M\"{u}nchen).

Born, Max (1971). {\it The Einstein Born Letters} 
(MacMillan, London) letter no.71.

Burke, William (1969). Ph.D. Thesis, Caltech.

Burke, William and Kip S. Thorne (1970). ``Gravitational Radiation Damping''
in {\it Relativity} eds. Moshe Carmeli, Stuart I. Fickler and Louis Witten
(Plenum Press, New York) pg. 209-228.

Cattani, Carlo and De Maria, Michelangelo (1993). ``Conservation Laws and
Gravitational Waves in General Relativity (1915-1918)'' in {\it The Attraction
of Gravitation: New Studies in the History of General Relativity} eds.
John Earman, Michel Janssen and John D. Norton 
(Birkh\"{a}user, Boston) pg. 63-87.

Chandrasekhar, Subrahmanyan (1965). 
``The post-Newtonian equations of Hydrodynamics
in General Relativity''
{\it Astrophysical Journal}, {\bf 142}, 1488-1512.

Chandrasekhar, Subrahmanyan and Esposito, F.P. (1970). ``The 
$2 {1\over 2}$-post-Newtonian
Equations of Hydrodynamics and Radiation Reaction in General Relativity''
{\it Astrophysical Journal}, {\bf 160}, 153-179.

Collins, Harry M. (1975). ``The Seven Sexes: A Study in the Sociology of
a Phenomenon, or the Replication of Experiments in Physics'' {\it Sociology},
{\bf 9}, 205-24.

Collins, Harry M. (1981). ``Son of Seven Sexes: The Social Destruction of a
Physical Phenomenon'' {\it Social Studies of Science}, {\bf 11}, 33-63.

Collins, Harry M. (1985). {\it Changing Order} (Sage Publications, London).

Cooperstock, Fred I. (1967). ``Energy Transfer via Gravitational 
Radiation in the Quasistellar Sources'', 
{\it Physical Review}, {\bf 163}, 1368-1373.

Cooperstock, Fred I. (1992) ``Energy Localization in General Relativity: A
New Hypothesis'', {\it Foundations of Physics}, {\bf 22}, 1011-1024.

Cooperstock, Fred I. and Hobill, D.W. (1982). ``Gravitational Radiation and
the Motion of Bodies in General Relativity''
{\it General Relativity and Gravitation}, {\bf 14}, 361-378. 

Damour, Thibault (1982). ``Gravitational Radiation and the Motion of Compact
Bodies'' in
{\it Rayonnement Gravitationelle} eds. N. Deruelle and T. Piran
(North Holland, Amsterdam) pg. 59-144.

Damour, Thibault (1983). ``Gravitational Radiation Reaction in the Binary
Pulsar and the Quadrupole-Formula Controversy'' {\it Physical Review Letters}
{\bf 51}, 1019-1021.

Damour, Thibault (1987a). ``An Introduction to the Theory of Gravitational
Radiation'' in {\it Gravitational in Astrophysics} eds. Brandon Carter and
James B. Hartle (Plenum Press, New York) pgs. 3-62.

Damour, Thibault (1987b). ``The problem of motion in Newtonian and
Einsteinian graivty'' in {\it 300 years of Gravitation} eds. Stephen
Hawking and Werner Israel (University Press, Cambridge) pgs. 128-198.

Damour, Thibault and Ruffini, R. (1974). ``Sur certaines v\'{e}rifications
nouvelles de la Relativit\'{e} g\'{e}n\'{e}rale rendues possibles par la
d\'{e}couverte d'un pulsar membre d'un syst\'{e}me binaire'' {\it Comptes 
Rendu de l'Academie des Sciences
de Paris, series A} {\bf 279}, 971-973.

De Witt, Cecile M. (1957). 
{\it Conference on the Role of Gravitation in Physics},
proceedings of conference at 
Chapel Hill, North Carolina, January 18-23, 1957. 
(Wright Air Development Center (WADC) technical
report 57-216, United States Air Force, 
Wright-Patterson Air Force Base, Ohio). A supplement with an expanded 
synopsis of Feynman's remarks was also distributed to participants (a copy
can be found, for example, in the Feynman papers at Caltech).

Dyson, Freeman (1963). ``Gravitational Machines'' in {\it Interstellar 
Communications} ed. A.G.W. Cameron (W.A. Benjamin Inc., New York) pg. 115-
120.

Eddington, Arthur Stanley (1922). ``The Propagation of Gravitational Waves''
{\it Proceedings of the Royal Society of London, series A}, {\bf 102}, 268-282. 

Ehlers, J\"{u}rgen (1979). {\it Isolated Gravitating systems in General
Relativity} Internation School of Physics, ``Enrico Fermi'' (North Holland,
Amsterdam).

Ehlers, J\"{u}rgen, Rosenblum, Arnold, Goldberg, Joshua N. 
and Havas, Peter (1976).
``Comments on Gravitational Radiation Damping and Energy Loss in
Binary Systems'', {\it The Astrophysical Journal}, {\bf 208}, L77-L81. 

Einstein, Albert (1916). ``N\"{a}herungsweise Integration der Feldgleichungen
der Gravitation''
{\it K\"{o}niglich Preussische Akademie der Wissenschaften Berlin,
Sitzungsberichte:} 688-696

Einstein, Albert (1918).  ``\"{U}ber Gravitationswellen''
{\it K\"{o}niglich Preussische Akademie der Wissenschaften Berlin,
Sitzungsberichte:} 154-167

Einstein, Albert and Grommer, Jakob (1927). 
``Allgemeine Relativit\"{a}tstheorie
und Bewegungsgesetz'' {\it Preussische Akademie der Wissenschaften Berlin,
Sitzungsberichte:} 2-13.

Einstein, Albert, Infeld, Leopold and Hoffmann, Banesh (1938). 
``The Gravitational Equations and the Problem of Motion'' {\it Annals of
Mathematics}, {\bf 39}, 65-100. 

Einstein, Albert and Ritz, Walter (1909). ``Zum gegenw\"{a}rtigen stand
des Strahlungsproblems'' {\it Physikalische Zeitschrift} {\bf 10}, 323-324.
Translated by Anna Beck in {\it Collected Papers of Albert Einstein Vol. 2,
The Swiss Years: Writings 1900-1909} (Princeton, Univ. Press) pg. 376.

Einstein, Albert and Rosen, Nathan (1937). ``On Gravitational Waves''
{\it Journal of the Franklin Institute}, {\bf 223}, 43-54.

Eisenstaedt, Jean (1986a). ``La relativit\'{e} g\'{e}n\'{e}rale a
l'\'{e}tiage: 1925-1955'' {\it Archive for the History of Exact Sciences} 
{\bf 35}, 115-185.

Eisenstaedt, Jean (1986b).``The Low water Mark of General Relativity,
1925-1955'' in
{\it Einstein and the History of General Relativity} eds. D. Howard and J. 
Stachel (Birkh\"{a}user, Boston) pg. 277-292.

Feynman, Richard P. and Leighton, Ralph (1988). 
{\it What do \underline{you} care what other people think? 
Further adventures of a curious character} 
(Norton, New York). Remark quoted appears
on pg. 91 of the Bantam paperback edition (New York, 1989). 

Fock, Vladimir A (1959). {\it Spacetime and Gravitation}, 
section 90 (Pergamon, New York) First English edition.

Fowler, William A. (1964). 
``Massive Stars, Relativistic Polytropes and Gravitational
Radiation: Gravitational Waves as Trigger for Radio Galaxy Emissions'' 
{\it Reviews of Modern Physics}, {\bf 36}, 545-555.

Franklin, Alan (1994). ``How to avoid the Experimenters' Regress'' {\it
Studies in the History and Philosophy of Science} {\bf 25}, 463-491.

Galison, Peter (1987). {\it How Experiments End} (University of Chicago Press).

Goldberg, Joshua N. (1955). ``Gravitational Radiation'' 
{\it Physical Review}, {\bf 99}, 1873-1883.

Goldberg, Joshua N. (1988). ``US Air Force Support of General Relativity:
1956-1972'' in {\it Studies in the History of General Relativity} eds.
Jean Eisenstaedt and A.J. Kox (Birkh\"{a}user, Boston).

Havas, Peter (1973). ``Equations of Motion, Radiation Reaction,
and Gravitational Radiation'' in {\it Ondes et Radiation Gravitationelles}
proceedings of meeting, Paris, June, 1973 (Editions du Centre National
de la recherche scientifique, Paris) pg. 383-392.

Havas, Peter (1979) ``Equations of Motion and Radiation Reaction in the
Special and General Theory of Relativity'' in {\it Isolated Gravitating Systems
in General Relativity} ed. J\"{u}rgen Ehlers (North Holland, Amsterdam) 
pg. 74-155.

Havas, Peter (1989). 
``The Early History of the Problem of Motion in General Relativity'' in 
{\it Einstein and the History of General Relativity} 
eds. D. Howard and J. Stachel
(Boston, Birkh\"{a}user) pg. 234-276.

Havas, Peter (1993). ``The Two-Body Problem and the Einstein-Silberstein
Controversy'' in {\it The Attraction of Gravitation} eds. John Earman,
Michel Janssen and John D. Norton (Birkh\"{a}user, Boston) pgs. 88-125.

Havas, Peter and Goldberg, Joshua N. (1962). ``Lorentz-Invariant Equations
of Motion of Point Masses in the General Theory of Relativity''
{\it Physical Review}, {\bf 128}, 398-414.

Hu, Ning (1947). ``Radiation Damping in the General Theory of Relativity''
{\it Proceedings of the Royal Irish Academy},
{\bf 51A}, 87-111.

Hulse, R.A. and Taylor, J.H. (1975). 
``Discovery of a Pulsar in a Binary System''
{\it Astrophysical Journal}, {\bf 195}, L51-L53.

Infeld, Leopold (1941). {\it Quest - The Evolution of a Physicist}
(Gollancz, London).

Infeld, Leopold and Plebanski, Jerzy (1960). {\it Motion and Relativity}
(Pergamon, New York).

Infeld, Leopold and Scheidegger, Adrian E. (1951). 
``Radiation and Gravitational
Equations of Motion'' {\it Canadian Journal of
Mathematics}, {\bf 3}, 195-207.

Infeld, Leopold and Wallace, Philip (1940). ``The Equations of Motion in
Electrodynamics''
{\it Physical Review}, {\bf 57}, 797-806.

Isaacson, Richard (1968). ``Gravitational Radiation in the limit of high
frequency II: Nonlinear terms and the Effective Stress Tensor'' 
{\it Physical Review} {\bf 166}, 1272-1280.

Jungnickel, Christa and McCormmach, Russel (1986). {\it Intellectual Mastery
of Nature, volume 2: The Now Mighty Theoretical Physics 1870-1925} 
(University of Chicago Press).

Kerr, Roy P. (1959). ``On the Lorentz-Invariant Approximation
Method in General Relativity III: The Einstein-Maxwell Field''
{\it Nuovo Cimento} {\bf 13}, 673-89.

Landau, Lev D. and Lifschitz, Evgenii M. (1951) {\it Classical Theory
of Fields} (Addison-Wesley, Cambridge, Mass).

Laplace, Pierre (1776). ``Sur le Principe de la Gravitation Universelle'' 
reprinted in
{\it Ouevres compl\'{e}tes de Laplace VIII} 
(Gauthier-Villars et fils, Paris, 1891)
pg. 201-275.

Laplace, Pierre (1825). {\it Trait\'{e} De M\'{e}canique C\'{e}leste}, Vol.4
Book X, Chapter VII, section 22 reprinted in {\it Ouevres compl\'{e}tes
de Laplace} (Gauthier-Villars et fils, Paris, 1891).

Levi-Civita, Tullio (1917). ``Sulla espressione analitica spettante al tensore
gravitzionale nella teorie di Einstein.'' {\it Rendiconti Accademia dei Lincei}
ser. 5, vol. XXVI, 381-391.

Lichnerowicz, Andr\'{e} (1955). {\it Theories relativistes de la gravitation
at de l'electromagnetisme} (Masson, Paris).

Lightman, Alan
and Brawer, Roberta (1990). 
{\it Origins} (Harvard University Press, Cambridge, Mass)

McVittie, G.C. (1955). ``Gravitational Waves and One-Dimensional Einsteinian
Gas Dynamics'' {\it Journal of Rational Mechanics and Analysis} {\bf 4}, 
201-220.

Pais, Abraham (1982). {\it Subtle is the Lord ... the Science and Life of
Albert Einstein} (Clarendon, Oxford) pgs. 494-495.

Peres, Asher (1959). ``Gravitational Motion and Radiation II''
{\it Nuovo Cimento} {\bf 11}, 644-655.

Peres, Asher (1960). ``Gravitational Radiation''
{\it Nuovo Cimento} {\bf 15}, 351-369.

Petrov, A. Z. (1955). {\it Doklady Akademii Nauk, SSSR}
{\bf 105}, 905.

Pirani, Felix A.E. (1955). Review of McVittie (1955) in {\it Mathematical
Reviews} {\bf 16}, 1165.

Poincar\'{e}, Henri (1908). ``La dynamique de l'electron'' 
{\it Revue g\'{e}n\'{e}rale des sciences pures et appliqu\'{e}s}
{\bf 19} 386-402 (1908), reprinted in {\it Oeuvres de Henri Poincar\'{e}}
vol. IX, pgs. 551-586 and 
translated by Francis Maitland as
``The New Mechanics'' in {\it Science and Method} (Dover, New York, 1952)
pgs. 199-250.

Ritz, Walter (1908).  {\it Critical Researches on General Electrodynamics},
``Recherches Critiques Sur l'Ectrodynamique Generale'' 
{\it Annales de Chemie et
de Physique}, {\bf 13}, 145-275.
Translation by R.S. Fitzius (self published by the translator, 1980).

Robertson, Howard Percy (1938). {\it Annals of Mathematics} {\bf 39}, 101.

Robinson, Ivor, Schild Alfred and Schucking, E.L. (1965). {\it Quasi-stellar
sources and gravitational collapse, including the proceedings of the First
Texas Symposium on Relativistic Astrophysics} (University of Chicago Press).

Rosen, Nathan (1937). ``Plane Polarized Waves in the General Theory of
Relativity.'' {\it Physikalische Zeitschrift der Sowjetunion} {\bf 12}, 366-372.

Rosen, Nathan (1955). ``On Cylindrical Gravitational Waves'' in 
{\it Jubilee of Relativity Theory} proceedings of the anniversary
conference at Bern, July 11-16, 1955 (Birkh\"{a}user-Verlag, Basel) pg.171-175.

Rosen, Nathan (1958). ``Energy and Momentum of Cylindrical Gravitational
Waves'', {\it Physical Review}, {\bf 110}, 291-292. 

Rosen, Nathan (1979). ``Does Gravitational
Radiation Exist'', {\it General Relativity and Gravitation}, {\bf 10}, 351-364. 

Rosen, Nathan and Virbhadra, K.S. (1993). 
``Energy and Momentum of Cylindrical Gravitational Waves.''
{\it General Relativity and Gravitation}, {\bf 25}, 429-433.

Sachs, Rainer K. (1962). ``Gravitational Waves in General Relativity VIII:
Waves in asymptotically flat space-time'' {\it Proceedings of the Royal
Society of London, series A} {\bf 270}, 103-126.

Scheidegger, Adrian E. (1951). ``Gravitational Transverse-Transverse
Waves'' {\it Physical Review}, {\bf 82}, 883-885.

Scheidegger, Adrian E. (1953). ``Gravitational Motion'' 
{\it Reviews of Modern Physics}, {\bf 25}, 451-468.

Schilpp, Paul Arthur (1949). {\it Albert Einstein, philosopher-scientist}
(Library of Living Philosophers, Evanston, Illinois).

Schwimming, Rainer (1980). ``On the History of the theoretical discovery
of the plane Gravitational Waves'' (preprint, Leipzig) 

Smith, Stanley F. and Havas, Peter (1965). ``Effects of Gravitational Radiation
Reaction in the General Relativistic two-body problem by a Lorentz-invariant
Method'' {\it Physical Review}, {\bf 138}, 495-508. 

Stachel, John (1959). ``Energy Flow in Cylindrical Gravitational Waves'',
Master's thesis (Stevens Institute of Technology).

Taylor, J.H. and McCulloch, P.M. (1980). 
``Evidence for the Existence of Gravitational
Radiation from measurements of the Binary Pulsar PSR 1913+16'' 
in {\it Proceedings of the 9th Texas symposium
on Relativistic Astrophysics}, 
eds. J\"{u}rgen Ehlers, Judith Perry and Martin Walker 
(New York Academy of Sciences, New York) pg. 442-446.

Tetrode, Hugo M. {\it Zeitschrift f\"{u}r Physik} {\bf 10}, 317-325. The
quoted translation is by J. Dorling.

Thorne, Kip S. (1980). ``Multipole Expansions of Gravitational Radiation''
{\it Reviews of Modern Physics} {\bf 52}, 299-339.

Thorne, Kip S. (1989). Unpublished manuscript on gravitational radiation, 
including a historical review of the radiation reaction controversy.

Trautman, Andrzej (1958a). ``Radiation and Boundary Conditions in the 
Theory of Gravitation'' {\it Bulletin de l'Academie 
Polonaise des Sciences, series
des Sciences mathematiques} {\bf 6}, 407-412.

Trautman, Andrzej (1958b). ``On Gravitational Radiation Damping'' {\it
Bulletin de l'Academie Polonaise des Sciences, series des Sciences
mathematiques} {\bf 6}, 627-633. 

Trautman, Andrzej, Pirani, Felix and Bondi, Hermann (1965). {\it Lectures
in General Relativity} (Prentice-Hall, New Jersey)

Walker, Martin and Will, Clifford M. (1980). ``The Approximation of Radiative
Effects in Relativistic Gravity: Gravitational Radiation Reaction and Energy
Loss in Nearly Newtonian Systems.''
{\it The Astrophysical Journal},
{\bf 242}, L129-L133.

Weber, Joseph and Wheeler, John Archibald (1957). ``Reality of the Cylindrical
Gravitational Waves of Einstein and Rosen''
{\it Reviews of Modern Physics}, {\bf 29}, 509-515.

Weber, Joseph (1960). ``Detection and Generation of Gravitational Waves''
{\it Physical  Review}, {\bf 117}, 306-313. 

Weber, Joseph (1969). ``Evidence for Discovery of Gravitational Radiation''
{\it Physical Review Letters}, {\bf 22}, 1320-1324.

Westpfahl, Konradin (1985). ``High speed scattering of charged and
and uncharged particles in General Relativity'' {\it Fortschritte
der Physik} {\bf 33}, 417-493.

Weyl, Hermann (1921). {\it Raum, Zeit, Materie} 4th ed. (Springer, Berlin)
pg. 228. English translation 
{\it Space, Time, Matter} (Metheun, London, 1922) pg. 252.

Wheeler, John Archibald and Feynman, Richard Phillips (1945). ``Interaction
with the Absorber as the Mechanism of Radiation'' {\it Reviews of Modern
Physics}, {\bf 17}, 157-181.

Wheeler, John Archibald and Feynman, Richard Phillips (1949). ``Classical
Electrodynamics in 
Terms of Direct Interparticle Action'' {\it Reviews of Modern
Physics} {\bf 21}, 425-433.

\end{document}